\documentclass[twocolumn,pra,aps,superscriptaddress]{revtex4}

\usepackage{xspace,amsmath,amsfonts,amsthm,amssymb,amsbsy}
\usepackage{graphicx}
\usepackage{amssymb}
\usepackage{color}
\usepackage{psfrag}
\usepackage{ifsym}
\usepackage{hyperref}

\usepackage{lipsum}

\definecolor{pink}{rgb}{1,0.078,0.57}

\definecolor{green}{rgb}{0,0.7,0.9}

\newcommand{\ket}[1] {| #1 \rangle}
\newcommand{\bra}[1] {\langle #1 |}

\newcommand{\dg}{^{\dagger}}

\begin{document}

\title{Thermal Light as a Mixture of Sets of Pulses: the Quasi-1D Example}

\author{Agata M. Bra\'nczyk}
\email{abranczyk@perimeterinstitute.ca}
\affiliation{Perimeter Institute for Theoretical Physics, Waterloo, Ontario, N2L 2Y5, Canada}
\author{Aur\'elia Chenu}
\affiliation{Department of Chemistry, Massachusetts Institute of Technology, Cambridge, Massachusetts 02139, USA}
\author{J. E. Sipe}
\affiliation{Department of Physics, 60 Saint George Street, University of Toronto, Toronto, Ontario, M5R 3C3 Canada}

\date{\today}

\begin{abstract}
The relationship between thermal light and coherent pulses is of fundamental and practical interest. We now know that thermal light cannot be represented as a statistical mixture of \emph{single} pulses. In this paper we ask whether or not thermal light can be represented as a statistical mixture of \emph{sets} of pulses. We consider thermal light in a one-dimensional wave-guide, and find a convex decomposition into products of orthonormal coherent states of localized, nonmonochromatic modes. 
\end{abstract}

\maketitle

\section{Introduction}

Until the invention of the laser, sources of visible light relied on thermal radiation. Lasers revolutionized medicine, science and technology, but thermal light continues to play an  important role in these fields. In quantum optics, thermal light has recently been used for novel types of imaging \cite{Gatti2004,Ferri2005,Valencia2005} and interferometry \cite{Chekhova1996}, and even as a resource for quantum mechanical protocols \cite{Guzman-Silva2016} and generation of non-classical states of light \cite{Bogdanov2016}. 

The relationship between thermal light and laser light was recently reviewed by Wiseman \cite{Wiseman2016} for continuous-wave (CW) lasers, and discussed by us  for broadband coherent pulses \cite{Chenu2015,Chenu2015b}. Decomposing thermal light into CW (i.e. completely delocalized) modes provides some conceptual challenges. For example, how should one think about a photosynthetic organism or a photovoltaic cell absorbing a plane-wave photon? A decomposition of thermal light into localized pulses would thus be desirable. We already showed that thermal light cannot be decomposed into a statistical mixture of \emph{single} pulses  \cite{Chenu2015}, but whether or not it can be decomposed into \emph{sets} of pulses, and if so what would be their nature, is an unsolved problem. 

Given the importance of thermal radiation, the study of different convex decompositions of thermal radiation is an interesting problem in and of itself. The well-known Glauber-Sudarshan $P$-representation \cite{Sudarshan1963,Glauber1963} has tremendously simplified many calculations in quantum optics. Decompositions into  (delocalized)  Schmidt-like coherent modes have also been investigated by Bobrov \emph{et al.} \cite{Bobrov2013}.
 
In this paper we consider thermal light propagating in a quasi-1D geometry, such as in an optical fiber. We find that it is possible to construct a convex decomposition of the thermal equilibrium density operator into products of orthonormal coherent states of localized, nonmonochromatic modes. These modes correspond to the scaling function for the Shannon (sinc) wavelet \cite{Mallat2008}. The coherent states in the decomposition are the quantum analogue of localized, coherent pulses of light. Our decomposition can be applied to thermal light over any frequency range, where the range determines the width of the pulses. 

We begin by quantizing the fields inside the waveguide in Sec. \ref{sec:quant}, then briefly review the Glauber-Sudarshan $P$-representation of thermal states in Sec. \ref{sec:del}, 
and write the density operator from an explicit mixture of sets of monochromatic states. The modes involved here being \emph{delocalized}, we then partition the thermal state density operator into portions of $k$-space, and introduce \emph{localized} modes in Sec. \ref{sec:nonmon}. In Sec. \ref{sec:sets}, we use these localized modes to decompose the density operator for thermal light. We further determine the fields in terms of these modes, and plot the field variation for typical pulse sets in Sec. \ref{sec:fields}. We conclude in Sec. \ref{sec:conc}.

\section{Waveguide fields}\label{sec:quant}

We consider light propagation in a quasi-1D geometry; that is, the propagation direction is taken as $z$ and any background index of refraction is then taken as a function of only $x$ and $y$, $n=n(x,y)$. We neglect any material dispersion in the index of refraction, but that could be easily included in the treatment \cite{Bhat2006}, as could a more general dependence of the index of refraction on position \cite{Bhat2001}, as in a photonic crystal structure.

We treat $\textbf{D}(\textbf{r})$ and $\textbf{B}(\textbf{r})$ as the fundamental field operators
\begin{align}
\begin{split}\label{eq:fields}
\mathbf{D}(\mathbf{r})={}&\sum_{m}\sqrt{\frac{\hbar\omega(k_{m})}{2L}}a_m\mathbf{d}_{k_{m}}(x,y)e^{ik_mz}+h.c.\,,\\
\mathbf{B}(\mathbf{r})={}&\sum_{m}\sqrt{\frac{\hbar\omega(k_{m})}{2L}}a_m\mathbf{b}_{k_{m}}(x,y)e^{ik_mz}+h.c.\,,
\end{split}
\end{align}
where $\mathbf{d}_{k_{m}}(x,y)$ and $\mathbf{b}_{k_{m}}(x,y)$ are appropriately normalized \cite{Bhat2006,Bhat2001}, with $k_m=2\pi m/L$, where $m$ is a nonzero integer and $L$ is the quantization length.  In  describing the fields, we have restricted ourselves to one mode type, since in thermal equilibrium the full density operator is a direct product of the density operators for the different mode types (in free space, the mode type could be polarization). We use the index $m$ to identify the lowering and raising operators, $a_{m}$ and $a_{m}^{\dagger}$ respectively, and $\omega(k)$ specifies the dispersion relation of the mode type of interest. Ignoring the zero-point energy, the Hamiltonian takes the form:
\begin{align*}
H=\sum_{m}\hbar\omega(k_m)a^{\dagger}_ma_m\,.
\end{align*}

\section{The thermal density operator and partitions of it}\label{sec:del}

We now look at the density operator for one of the modes $m$. There are many ways to write out the density operator of a harmonic oscillator in thermal equilibrium; here we will consider the convex decomposition in terms of coherent states,
\begin{align}\label{eq:alpha}
\ket{\alpha_m}=e^{\alpha_ma^{\dagger}_m-\alpha^*_ma_m}\ket{vac}\,,
\end{align}
where $\alpha_m$ is a complex number. 

The density operator for thermal equilibrium of mode $m$ can be written as
 \begin{align}\label{eq:rhom}
\rho_{m}={}&\int \frac{d^2\alpha_m}{\pi \langle n_m \rangle}e^{-\frac{~|\alpha_m|^2}{\langle n_m \rangle}}\ket{\alpha_m}\bra{\alpha_m}\,,
\end{align}
where $\langle n_m \rangle=(e^{\beta \hbar\omega(k_m)}-1)^{-1}$. Equation (\ref{eq:rhom}) is the Glauber-Sudarshan $P$-representation \cite{Sudarshan1963,Glauber1963} for thermal light in mode $m$.

It is not always necessary to consider the entire spectrum of thermal radiation. For some applications, only part of the spectrum may be relevant. It is therefore useful to partition $\rho$ as
\begin{align*}
\rho=\prod_{J}^{\otimes}\rho_{J}\,,
\end{align*}
where $\rho_J$ is the density operator associated with a portion $J$ of $k$-space:
\begin{align}\label{eq:J}
\rho_{J}=\prod_{m\in S}^{\otimes}\rho_{m}\,.
\end{align}
The tensor product is over a finite set of discrete modes with wavenumbers $\{k_m\}$, defined by the finite set of consecutive integers  $S$. The union of all $S$ is $\mathbb{Z}$. 

\begin{figure}[t]
\includegraphics[width=\columnwidth]{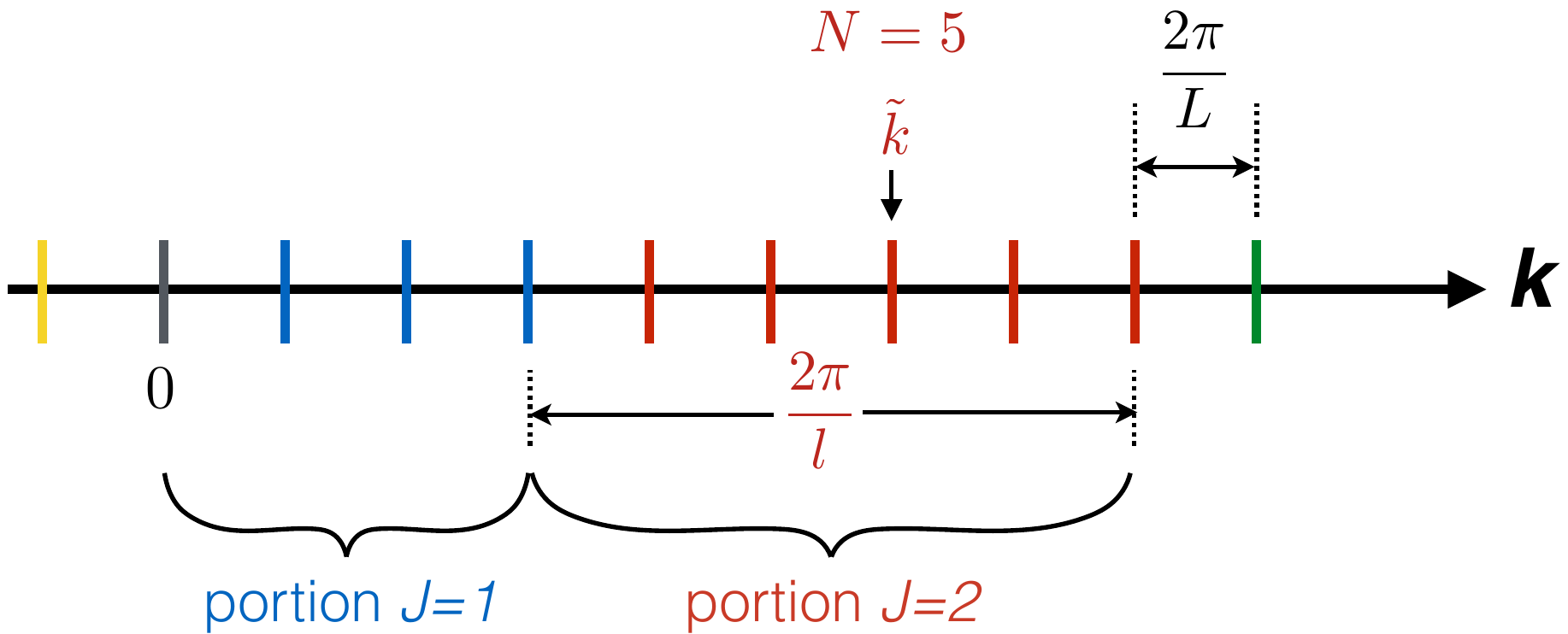}
\caption{Schematic of the discrete modes of the waveguide. Each colour corresponds to a different portion $J$ of $k$-space, which defines the partition $\rho_J$ of the thermal density operator.   \label{fig:2}}
\end{figure}

For convenience, we take the number of modes $N$ in each partition to be odd, and write  $\{k_m\}=\{\tilde k+\kappa_{m}\}_{m=-n,\dots,n}$, where $n=(N-1)/2$. We also introduce a lattice constant, defined as $l=L/N$, which defines the region of $k$-space in portion $J$. Fig. \ref{fig:2} shows a schematic of the modes. Each color corresponds to a different set $J$. 

In this notation, the partition 
 \begin{align}\label{eq:rhoset}
\rho_{J}={}&\int  \left(\prod_{m=-n }^{n}\frac{d^2\alpha_m}{\pi \langle n_m \rangle}\right)e^{-\sum_{m}\frac{~|\alpha_m|^2}{\langle n_m \rangle}}\ket{\{\alpha\}}\bra{\{\alpha\}}\,,
\end{align}
where $\langle n_m \rangle=(e^{\beta \hbar\omega(\tilde{k}+\kappa_m)}-1)^{-1}$. Because the operators $a_{m}$ satisfy the canonical commutation relations, a set of monochromatic coherent states can be defined as $\ket{\{\alpha\}}=\prod_{m}^{\otimes}\ket{\alpha_{m}}$, where $m=-{n},\dots, {n}$. 

For the remainder of this paper, we will consider thermal light within a particular portion $J$. 

\section{Nonmonochromatic modes and localized pulses}\label{sec:nonmon}

The operators $a_{m}$ and $a_{m}^{\dagger}$ introduced above are associated with modes that are delocalized over the length of the waveguide, and are characterized by eigenfrequencies. In this section we introduce more general, nonmonochromatic modes \cite{Titulaer1966}, the coherent states of which describe localized pulses.

Nonmonochromatic modes can be created by making a canonical transformation, introducing new lowering and raising operators ${c}_{s}$ and ${c}^{\dagger}_{s}$. We write $c_s\equiv \sum_{m=-n}^{n}C_{sm}a_m$. The operators satisfy $[{c}_{s},{c}_{s'}^{\dagger}]=\delta_{ss'}$. These modes can be used to build nonmonochromatic coherent states
\begin{align*}
\ket{\gamma_{s}}=e^{\gamma_{s}c_{s}^{\dagger}-\gamma^*_{s}c_{s}}\ket{vac}\,,
\end{align*}
where $\gamma_s$ is a complex number.

Because the operators $c_{s}$ also satisfy the canonical commutation relations, a set of nonmonochromatic coherent states can be defined as $\ket{\{\gamma\}}=\prod_{s}^{\otimes}\ket{\gamma_{s}}$, where $s=-{n},\dots, {n}$. 

To define modes that correspond to localized pulses when excited in a coherent state, we first introduce a set of wave functions $\phi_{m}(z)=e^{i\tilde kz}\chi_{m}(z)$ where $\chi_{m}(z)\equiv e^{i\kappa_mz}/\sqrt{L}$. The wave functions are orthonormal over  the quantization length $L$, such that $\int_{-L/2}^{L/2}\phi^*_{m}(z)\phi_{m'}(z)=\delta_{mm'}$. From $\chi_{m}(z)$,  we construct a function 
\begin{align*}
w(z)={}\frac{1}{\sqrt{N}}\sum_{m=-{n}}^{n}\chi_{m}(z)={}\frac{1}{\sqrt{NL}}\frac{\sin(\frac{\pi z}{l})}{\sin(\frac{\pi z}{L})}\,.
\end{align*}

Since $\chi_{m}(z+L)=\chi_{m}(z)$ we also have that $w(z+L)=w(z)$; that is, it is periodic over the periodic length $L$. Nonetheless, for $z$ close to zero $w(z)$ initially drops off like a sinc function. 

We now introduce a set of associated functions $w_{s}(z)\equiv w(z-sl)$. The functions are also orthonormal over  $L$, such that $\int_{-L/2}^{L/2}w^*_{s}(z)w_{s'}(z)dz=\delta_{ss'}$. 

The relationship between the basis functions can be written in the form
\begin{align*}
\chi_m(z)={}&\sum_{s=-{n}}^{n}w_{s}(z)C_{sm}\,,\\
w_{s}(z)={}&\sum_{m=-{n}}^{n}\chi_m(z)C^*_{sm}\,,
\end{align*}
where
\begin{align*}
C_{sm}=\frac{1}{\sqrt{N}}e^{\frac{2\pi i s l}{L}m}\,.
\end{align*}

The functions $w_s(z)$ will be associated with the operators ${c}_{s}$ and ${c}_{s}^{\dagger}$. In the limit $L\rightarrow\infty$,  $w_s(z)$ will become localized, and will correspond to the scaling function for the Shannon (sinc) wavelet.

In this limit, the connection between the nonmonochromatic coherent state $\ket{\gamma_s}$ and a ``classical'' coherent pulse can be made by taking the expectation value of the displacement field operator with respect to the nonmonochromatic coherent state, i.e., $\langle\mathbf{D}(\mathbf{r})\rangle=\bra{\gamma_s}\mathbf{D}(\mathbf{r})\ket{\gamma_s}$. We do this in Sec. \ref{sec:fields} for a particularly simple limiting case.

\section{Thermal light as a mixture of sets of pulses}\label{sec:sets}
We can now proceed to write the density operator of thermal light as a mixture of   \emph{sets} of localized pulses. We begin with $\rho_J$ as a mixture  of sets of monochromatic modes, as in Eq. (\ref{eq:rhoset}). We then put $\gamma_{s}\equiv\sum_{m=-n}^{n}C_{sm}\alpha_m$, and use the definition $c_s\equiv \sum_{m=-n}^{n}C_{sm}a_m$,  such that
\begin{align*}
\ket{\{\alpha\}}=\exp\left(\sum_{s=-n}^{n}\gamma_sc^{\dagger}_s-\gamma^*_sc_s\right)\ket{vac}=\ket{\{\gamma\}}\,.
\end{align*}
Changing variables $\gamma_s=\bar\gamma_se^{-\Gamma}$,  where 
\begin{align*}
\Gamma=\frac{1}{2N}\sum_{m=-n}^{n}\ln(e^{\beta\hbar\omega(\tilde k+\kappa_m)}-1)\,,
\end{align*}
the decomposition in Eq. (\ref{eq:rhoset}) is then rewritten as
\begin{align}\label{eq:rhoJ3}
\rho_{J}=\int \left(\prod_{s=-n}^{n} \frac{d^2\bar\gamma_{s}}{\pi}\right) F(\{\bar\gamma\})\ket{\{\bar\gamma e^{-\Gamma}\}}\bra{\{\bar\gamma e^{-\Gamma}\}}\,,
\end{align}
where since a unitary matrix relates the $\gamma_s$ to the $\alpha_m$, we have taken $\prod_m d^2\alpha_m=\prod_s d^2\gamma_s$. 
The probability density function is
\begin{align*}
F(\{\bar\gamma\})={}&\exp\left(-\sum_{s=-n}^{n}\sum_{s'=-n}^{n}\bar\gamma_{s}\Lambda_{ss'}\bar\gamma^*_{s'}\right)\,,
\end{align*}
where
\begin{align*}
\Lambda_{ss'}=e^{-2\Gamma}\sum_{m=-n}^{n}C^*_{sm}C_{s'm}(e^{\beta\hbar\omega(\tilde k+\kappa_m)}-1)\,.
\end{align*}

\begin{figure*}[t]
\begin{flushleft}
a)\hspace{3.85cm}b)\hspace{7.15cm}c)\\
\includegraphics[width=\textwidth]{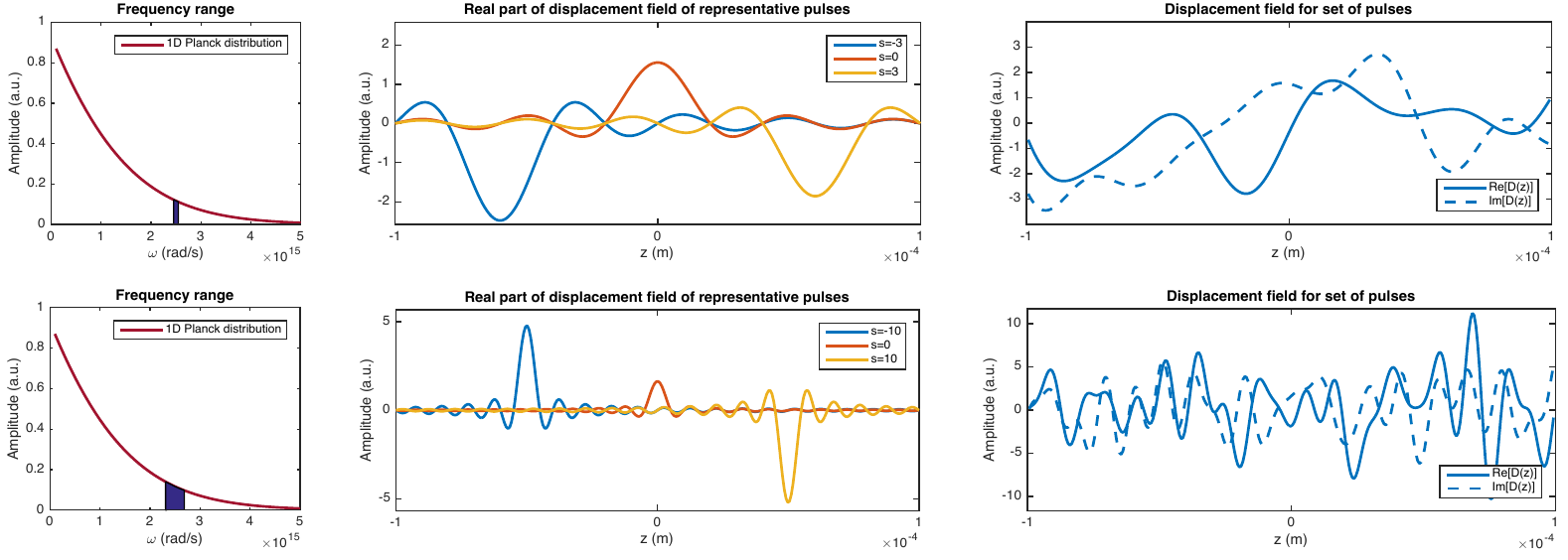}
\caption{Mean displacement fields for $\ket{\{\bar\gamma e^{-\Gamma}\}}$ spanning two representative portions $J$ of  $k$-space; one narrower (top) and one wider (bottom). In each case, we consider a typical pulse set, as defined in Appendix \ref{sec:typ}. The shaded region in Figure a) indicates the $k$-range with respect to the 1D Planck distribution. Figure b) shows the displacement field for selected individual pulses, given by the real part of the first term of Eq. (\ref{eq:Ds}), evaluated at $x=y=0$, with the carrier wave $e^{i\tilde k z}$ omitted for image clarity. Figure c) shows the total displacement field for a set of pulses, given by the first term of Eq. (\ref{eq:D}), evaluated at $x=y=0$, with the carrier wave $e^{i\tilde k z}$ omitted for image clarity.  }
\label{fig:1}
\end{flushleft}
\end{figure*}

The density matrix in (\ref{eq:rhoJ3}) was derived from (\ref{eq:rhoset}) and is therefore equivalent to it. As all correlation functions are determined by the density matrix \cite{Glauber1963,Mandel1995}, any correlation function calculated from the mixture of sets of pulses (\ref{eq:rhoJ3}) will be equivalent to the familiar correlation functions calculated from the standard expression (\ref{eq:rhoset}).

We now take the limit to infinite normalization length, as detailed in Appendix \ref{app:lim}. The range of $s$ in the product and summations appearing above goes to $-\infty$ to $\infty$, and $s$ ranges over all the integers. The range of $\kappa_m$ within portion $J$ becomes $-\pi/l<\kappa\leq \pi/l$, where $ \kappa$  indicates the continuous version of $\kappa_{m}$ as $L\rightarrow\infty$.

We arrive at the main result of this paper, that is, thermal light, in portion  $J$ of $k$-space, decomposed into states $\ket{\{\bar\gamma e^{-\Gamma}\}}$, with probability density function $F(\{\bar\gamma\})$:
\begin{align}\label{eq:rhoJinf}
\rho_{J}=\int \left(\prod_{s\in\mathbb{Z}} \frac{d^2\bar\gamma_{s}}{\pi}\right) F(\{\bar\gamma\})\ket{\{\bar\gamma e^{-\Gamma}\}}\bra{\{\bar\gamma e^{-\Gamma}\}}\,,
\end{align}
 where now 
\begin{align*}
F(\{\bar\gamma\})={}&\exp\left(-\sum_{s\in\mathbb{Z}}\sum_{s'\in\mathbb{Z}}\bar\gamma_{s}\Lambda_{ss'}\bar\gamma^*_{s'}\right)\,,
\end{align*} 
and 
 
\begin{align*}
\Gamma=\frac{l}{2}\int_{-\frac{\pi}{l}}^{\frac{\pi}{l}}\frac{d \kappa}{2\pi}\ln(e^{\beta\hbar\omega(\tilde k+\kappa)}-1)\,,
\end{align*}
and
\begin{align*}
\Lambda_{ss'}=e^{-2\Gamma}\int_{-\frac{\pi}{l}}^{\frac{\pi}{l}}\frac{d \kappa}{2\pi}e^{i(s-s') \kappa l}(e^{\beta\hbar\omega(\tilde k+\kappa)}-1)\,.
\end{align*}
The state $\ket{\{\bar\gamma e^{-\Gamma}\}}$ is a product of orthonormal coherent states of localized, nonmonochromatic modes of the waveguide. The state is the quantum analogue of a set of localized, coherent pulses of light. 

\section{The fields}\label{sec:fields}

We now write the fields in terms of the localized modes derived in Sec. \ref {sec:nonmon}.  This can simplify the calculation of certain quantities, such as field expectation values, associated with those of our sets of pulses. 

From Eq. (\ref{eq:fields}), the field for portion $J$ is
\begin{align*}
\begin{split}
\mathbf{D}(\mathbf{r})={}&\sum_{m=-n}^{n}\sqrt{\frac{\hbar\omega(\tilde k+\kappa_{m})}{2}}a_m\mathbf{d}_{\tilde k+\kappa_{m}}(x,y)\phi_{m}(z)+h.c.\,,\\
\mathbf{B}(\mathbf{r})={}&\sum_{m=-n}^{n}\sqrt{\frac{\hbar\omega(\tilde k+\kappa_{m})}{2}}a_m\mathbf{b}_{\tilde k+\kappa_{m}}(x,y)\phi_{m}(z)+h.c.\,.
\end{split}
\end{align*}

Making the canonical transformation and taking the normalization length $L$ to infinity, we get
\begin{align*}
\begin{split}
\mathbf{D}(\mathbf{r})={}&\sum_{s\in\mathbb{Z}}c_s\mathbf{d}(x,y;z-sl)e^{i\tilde kz}+h.c.\,,\\
\mathbf{B}(\mathbf{r})={}&\sum_{s\in\mathbb{Z}}c_s\mathbf{b}(x,y;z-sl)e^{i\tilde kz}+h.c.\,,
\end{split}
\end{align*}
where each term in the summation represents the field for a different localized pulse, centered at $sl$. The field  mode for each pulse is given by
\begin{align*}
\begin{split}
\mathbf{d}(x,y;z)={}&\sqrt{l}\int_{-\frac{\pi}{l}}^{\frac{\pi}{l}}\frac{d \kappa}{2\pi}\sqrt{\frac{\hbar\omega(\tilde k+\kappa)}{2}}\mathbf{d}_{\tilde k+\kappa}(x,y)e^{i\kappa z}+h.c.\,,\\
\mathbf{b}(x,y;z)={}&\sqrt{l}\int_{-\frac{\pi}{l}}^{\frac{\pi}{l}}\frac{d \kappa}{2\pi}\sqrt{\frac{\hbar\omega(\tilde k+\kappa)}{2}}\mathbf{b}_{\tilde k+\kappa}(x,y)e^{i\kappa z}+h.c.\,.\\
\end{split}
\end{align*}
where $\mathbf{d}_{\tilde k+\kappa}(x,y)$ and $\mathbf{b}_{\tilde k+\kappa}(x,y)$ are the transverse modes of the waveguide. 

For the purpose of illustration, consider a particularly simple limit, when the range of integration over $\kappa$ is taken to be so small that we can approximate $\omega(\tilde k+\kappa)$ as $\omega(\tilde k)$, and we can take $\mathbf{d}_{\tilde k+\kappa}(x,y)$ to be $\mathbf{d}_{\tilde k}(x,y)$ (and similarly for $\mathbf{b}_{\tilde k+\kappa}(x,y)$). In this limit, we have

\begin{align*}
\mathbf{d}(x,y;z)\approx{}&\sqrt{\frac{\hbar\omega(\tilde k)}{2}}\mathbf{d}_{\tilde k}(x,y)W(z)+h.c.\,,
\end{align*}
where 
\begin{align*}
W(z)=\frac{2\pi}{\sqrt{l}}\mathrm{sinc}\left(\frac{\pi z}{l}\right)\,.
\end{align*}
This can be understood as the limit of $w(z)$ as the normalization length goes to infinity; the function is thus now no longer periodic. In the language of solid state physics, this would be an ``empty lattice Wannier function.'' In analogy with $w_{s}(z)$ we can define $W_{s}(z)$ centered at different ``lattice sites'',
\begin{align*}
W_{s}=W(z-sl)\,.
\end{align*}
In this approximation, we can write
\begin{align*}
\begin{split}
\mathbf{D}(\mathbf{r})={}&e^{i\tilde kz}\sqrt{\frac{\hbar\omega(\tilde k)}{2}}\mathbf{d}_{\tilde k}(x,y)\sum_{s\in\mathbb{Z}}c_sW_{s}(z)+h.c.\,,\\
\mathbf{B}(\mathbf{r})={}&e^{i\tilde kz}\sqrt{\frac{\hbar\omega(\tilde k)}{2}}\mathbf{b}_{\tilde k}(x,y)\sum_{s\in\mathbb{Z}}c_sW_{s}(z)+h.c.\,.\\
\end{split}
\end{align*}
The expectation values for the fields of a pulse set $\ket{\{\bar\gamma e^{-\Gamma}\}}$ are
\begin{align}\label{eq:D}
\langle\mathbf{D}(\mathbf{r})\rangle={}&e^{i\tilde kz}\sqrt{\frac{\hbar\omega(\tilde k)}{2}}\mathbf{d}_{\tilde k}(x,y)\sum_{s\in\mathbb{Z}}\bar\gamma_se^{-\Gamma}W_{s}(z)+c.c.\,,\\\nonumber
\langle\mathbf{B}(\mathbf{r})\rangle={}&e^{i\tilde kz}\sqrt{\frac{\hbar\omega(\tilde k)}{2}}\mathbf{b}_{\tilde k}(x,y)\sum_{s\in\mathbb{Z}}\bar\gamma_se^{-\Gamma}W_{s}(z)+c.c.\,.
\end{align}
The expectation values for the fields of a single pulse $\ket{\bar\gamma_se^{-\Gamma}}$ are 
\begin{align}\label{eq:Ds}
\langle\mathbf{D}_s(\mathbf{r})\rangle={}&e^{i\tilde kz}\sqrt{\frac{\hbar\omega(\tilde k)}{2}}\mathbf{d}_{\tilde k}(x,y)\bar\gamma_se^{-\Gamma}W_{s}(z)+c.c.\,,\\\nonumber
\langle\mathbf{B}_s(\mathbf{r})\rangle={}&e^{i\tilde kz}\sqrt{\frac{\hbar\omega(\tilde k)}{2}}\mathbf{b}_{\tilde k}(x,y)\bar\gamma_se^{-\Gamma}W_{s}(z)+c.c.\,.
\end{align}

In Fig. \ref{fig:1} we show the positive frequency components of the mean displacement fields for typical pulse sets, as defined in Appendix \ref{sec:typ}. The fields for individual pulses are plotted in \ref{fig:1} b) and the field for a set of pulses is plotted in \ref{fig:1} c). Notice that the width of the pulses decreases as $\rho_J$ spans an increasingly larger range of $k$-space, as shown in  \ref{fig:1} a).

The mean fields of our pulse sets are very different from those of a pulse train generated by, say, a mode-locked laser. The latter has periodic bursts of high field amplitude interspaced by typically longer segments of zero field amplitude, whereas the  field amplitude of a typical pulse set described here is more complicated, as can be seen in Fig. \ref{fig:1} c).

\section{Conclusion}\label{sec:conc}

The relationship between thermal light and coherent pulses is of fundamental and practical interest. While thermal light cannot be represented as a statistical mixture of single localized pulses \cite{Chenu2015}, we have shown here how to decompose thermal light in a 1D waveguide into a statistical mixture of \emph{sets} of localized pulses. Our results can also be applied when there is no optical fiber or confining geometry, but rather there exists a ``column'' of light, neglecting diffraction. We plan to turn to more generalizations in later communications.

The form of the convex decomposition we have introduced makes modelling a finite frequency range very natural, while maintaining a representation in terms of localized pulses; this would arise when dealing with filtered thermal light. The decomposition also lends itself to treating thermal light that has been chopped in the spatial domain, as long as the length is much greater than the width of the function $w_s(z)$. This can simply be done by truncating the range of $s$ in Eq.   (\ref{eq:rhoJinf}). 

We emphasize that the decomposition presented here produces the thermal equilibrium density operator in 1D. All properties of thermal light in 1D such as correlation functions, ergodicity, stationarity, etc., will,  therefore, also be reproduced by this decomposition. Furthermore, all correlation functions will be independent of the properties of the pulses in the statistical mixture. We anticipate that this decomposition will serve as a useful tool for studying interactions of matter with thermal light in 1D.

\section{Acknowledgements}

AMB and AC thank the Kavli Institute for Theoretical Physics for hosting them during the Many-Body Physics with Light program. Research at Perimeter Institute is supported by the Government of Canada through Industry Canada and by the Province of Ontario through the Ministry of Research and Innovation. This research was supported in part by the National Science Foundation under Grant No. NSF PHY11-25915. AC acknowledges funding from the Swiss National Science Foundation and thanks P. Brumer and J. Cao for hosting her. JES acknowledges support from the Natural Sciences and Engineering Research Council of Canada. 

\appendix

\section{Taking the limit}\label{app:lim}

Consider a waveguide with a normalization length $L$, with $N$ modes in the portion $J$ of $k$-space. The modes are represented by the set $\{k_m\}$ where $m$ is taken from a set of consecutive integers.

\begin{figure}[t!]
\includegraphics[width=\columnwidth]{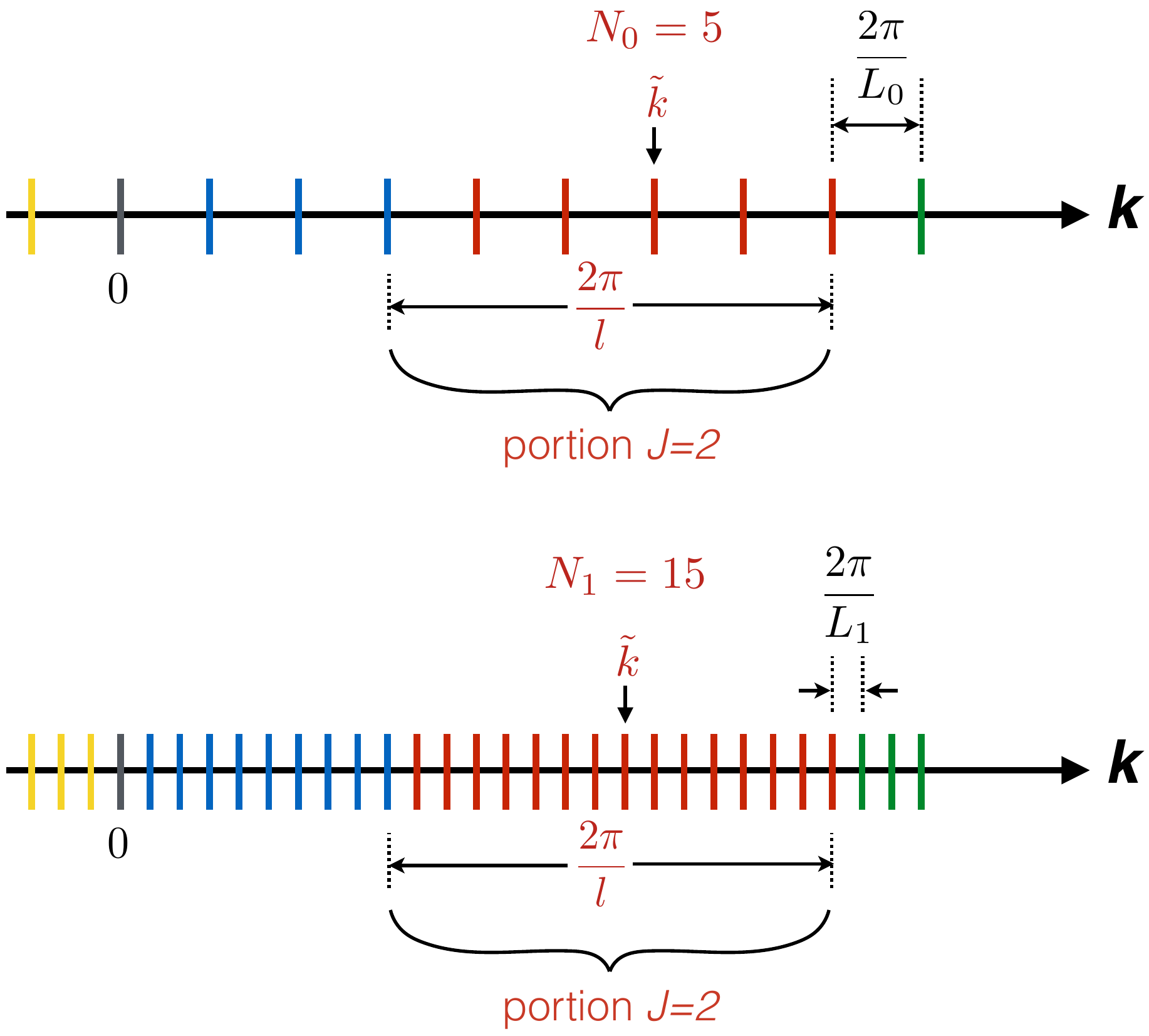}
\caption{Schematic of the discrete modes of the waveguide for $j=0$ (top) and $j=1$ (bottom). Each colour corresponds to a different portion $J$ of $k$-space. Note that as $j$ increases, $\tilde k$ moves closer to the centre of the region of length $2\pi/l$, where $l=L/N$ is the lattice constant. \label{fig:3}}
\end{figure}

But recall that $k_{m}=2\pi m/L$. As the normalization length increases, the density of modes also increases, and naively $\{k_m\}$ would span a progressively narrower portion of $k$-space. To take the limit to infinite normalization length while keeping fixed the range in $k$-space spanned by $\rho_J$, the number of elements in $\{k_m\}$ should be increased appropriately. We arrange this as follows. We relabel $L\equiv L_0$ and $N\equiv N_0$, and define  $L_j=3L_{j-1}$ and $N_j=3N_{j-1}$. The limit to infinite length is then acquired by taking $j\rightarrow\infty$. In this way, the lattice constant $l=L_j/N_j$ remains constant, as does the region on $k$-space. This is demonstrated schematically in Fig. \ref{fig:3}.

\section{Choosing typical pulse trains}\label{sec:typ}

Eq. (\ref{eq:rhoJinf}) is the expression for thermal light, in portion $J$ of $k$-space, decomposed into pulse sets. Each pulse set contains an infinite number of pulses. 

Now consider the field in a finite region of space; pulses far away from that region will make a negligible contribution. We therefore only consider a finite subset of all pulses when plotting the field variation. We denote this subset $\mathcal{S}$. Then what we really want to think about is
\begin{align*}
\rho_{\mathcal{S}}=\int \left(\prod_{s\in\mathcal{S}} \frac{d^2\bar\gamma_{s}}{\pi}\right) F(\{\bar\gamma\})\ket{\{\bar\gamma e^{-\Gamma}\}}\bra{\{\bar\gamma e^{-\Gamma}\}}\,,
\end{align*}
where
\begin{align}\label{eq:F2}
F(\{\bar\gamma\})={}&\exp\left(-\sum_{s\in\mathcal{S}}\sum_{s'\in\mathcal{S}}\bar\gamma_{s}\Lambda_{ss'}\bar\gamma^*_{s'}\right)\,,
\end{align} 
and 
\begin{align*}
\ket{\{\bar\gamma e^{-\Gamma}\}}=\prod^{\otimes}_{s\in\mathcal{S}}\ket{\bar\gamma_se^{-\Gamma}}\,.
\end{align*}

To plot the field variation for a ``typical'' pulse set, we want to select a likely  set $\bar\gamma_s e^{-\Gamma}$ from the distribution $F(\{\bar\gamma e^{-\Gamma}\})$, Eq. (\ref{eq:F2}). To do this, our approach is to write $F(\{\bar\gamma e^{-\Gamma}\})$ as as product of simpler distributions. Notice that the matrix $\Lambda_{ss'}$ is Hermitian, so it can be diagonalized by a unitary transformation. Then we have
\begin{align*}
\sum_{s,s'\in\mathcal{S}}U^*_{sr}\Lambda_{ss'}U_{s'r'}=\theta_{r}\delta_{rr'}{}\,,
\end{align*}
where $U\dg\Lambda U$ is a diagonal matrix with elements $\theta_r$. We put $\eta_r=\sum_{s\in\mathcal{S}}U_{sr}\bar\gamma_s$ and write
\begin{align*}
\rho= \int\left(\prod_{r\in\mathcal{S}}\frac{d^2\eta_{r}}{\pi}\right) e^{-\sum_{r\in\mathcal{S}}\theta_{r}|\eta_r|^2}\ket{\{\eta_r e^{-\Gamma}\}}\bra{\{\eta_r e^{-\Gamma}\}}\,.
\end{align*} 

We can now ask: In this mixture, how do we characterize the probability associated with the pulse set $\ket{\{\bar\gamma e^{-\Gamma}\}}$? 

Writing $\eta_r=|\eta_r|e^{i\phi_r}$, we have $d^2\eta_{r}=(d\phi_rd|\eta_r|)|\eta_r|$.  For each $r$, any $\phi_r$ is between  $0$ and $2\pi$ and is equally likely. But $|\eta_r|$, which ranges from $0$ to $\infty$, needs to be taken from the distribution $|\eta_r|e^{-\theta_r|\eta_r|}$, which peaks at $|\eta_r|=1/\sqrt{2\theta_r}$.

Our method for identifying very ``likely'' pulse sets is as follows. Find the matrix $U$ and  diagonal values  $\theta_r$  by diagonalizing $\Lambda$. Then for each $r$: 1) choose $\phi_r$ at random from $0$ to $2\pi$; and 2) take $|\eta_r|=1/\sqrt{2\theta_r}$. From the set of complex numbers $\eta_r$, and  the matrix  $U$, identify the set of  amplitudes $\bar\gamma_s=\sum_{r\in\mathcal{S}}U^*_{sr}\eta_r$.

For each random set of phases $\{\phi_r\}$, with each $|\eta_r|=1/\sqrt{2\theta_r}$, we will get a very ``likely'' pulse set, and those sets with different random phases will be ``equally likely''.  Pulse sets that are ``less likely'' can be  investigating by putting  $|\eta_r|\neq1/\sqrt{2\theta_r}$.

\end{document}